\documentclass[5p, number, final]{elsarticle}

\usepackage{amsmath}
\usepackage{graphicx}

\usepackage{hyphenat}
\journal{Physics Letters B}

\begin{document}

\begin{abstract}
We present the results of an experiment to search for trapped antihydrogen atoms with the ALPHA antihydrogen trap at the CERN Antiproton Decelerator.
Sensitive diagnostics of the temperatures, sizes, and densities of the trapped antiproton and positron plasmas have been developed, which in turn permitted development of techniques to precisely and reproducibly control the initial experimental parameters. 
The use of a position-sensitive annihilation vertex detector, together with the capability of controllably quenching the superconducting magnetic minimum trap, enabled us to carry out a high-sensitivity and low-background search for trapped synthesised antihydrogen atoms.
We aim to identify the annihilations of antihydrogen atoms held for at least $130~\mathrm{ms}$ in the trap before being released over $\sim 30~\mathrm{ms}$.
After a three-week experimental run in 2009 involving mixing of $10^7$ antiprotons with $1.3 \times 10^9$ positrons to produce $6 \times 10^5$ antihydrogen atoms, we have identified six antiproton annihilation events that are consistent with the release of trapped antihydrogen. 
The cosmic ray background, estimated to contribute 0.14 counts, is incompatible with this observation at a significance of 5.6 sigma. 
Extensive simulations predict that an alternative source of annihilations, the escape of mirror-trapped antiprotons, is highly unlikely, though this possibility has not yet been ruled out experimentally.
\end{abstract}


\title{Search for Trapped Antihydrogen\\[1cm] \normalsize{ALPHA Collaboration}}

\author[aarhus]{G.B.~Andresen}
\author[simonfraser]{M.D.~Ashkezari}
\author[berkeley]{M.~Baquero-Ruiz}
\author[swansea]{W.~Bertsche}
\author[aarhus]{P.D.~Bowe}
\author[berkeley]{C.C.~Bray}
\author[swansea]{E.~Butler}
\author[rio]{C.L.~Cesar}
\author[berkeley]{S.~Chapman}
\author[swansea]{M.~Charlton}
\author[berkeley]{J.~Fajans}
\author[calgary]{T.~Friesen}
\author[triumf,calgary]{M.C.~Fujiwara}
\author[triumf]{D.R.~Gill}
\author[aarhus]{J.S.~Hangst}
\author[ubc]{W.N.~Hardy}
\author[tokyo]{R.S.~Hayano}
\author[simonfraser]{M.E.~Hayden}
\author[swansea]{A.J.~Humphries}
\author[calgary]{R.~Hydomako}
\author[swansea,stockholm]{S.~Jonsell}
\author[swansea]{L.V.~J\o rgensen}
\author[triumf]{L.~Kurchaninov}
\author[rio]{R.~Lambo}
\author[swansea]{N.~Madsen}
\author[york]{S.~Menary}
\author[liverpool]{P.~Nolan}
\author[triumf]{K.~Olchanski}
\author[triumf]{A.~Olin}
\author[berkeley]{A.~Povilus}
\author[liverpool]{P.~Pusa}
\author[auburn]{F.~Robicheaux}
\author[negev]{E.~Sarid}
\author[ubc]{S.~Seif El Nasr}
\author[riken,tokyo2]{D.M.~Silveira}
\author[berkeley]{C.~So}
\author[triumf]{J.W.~Storey}
\author[calgary]{R.I.~Thompson}
\author[swansea]{D.P.~van der Werf}
\author[swansea]{D.~Wilding}
\author[berkeley]{J.S.~Wurtele}
\author[riken,tokyo2]{Y.~Yamazaki}

\address[aarhus]{Department of Physics and Astronomy, Aarhus University, DK-8000 Aarhus C, Denmark}
\address[simonfraser]{Department of Physics, Simon Fraser University, Burnaby BC, V5A 1S6, Canada}
\address[berkeley]{Department of Physics, University of California, Berkeley, CA 94720-7300, USA}
\address[swansea]{Department of Physics, Swansea University, Swansea SA2 8PP, United Kingdom}
\address[rio]{Instituto de F\'{i}sica, Universidade Federal do Rio de Janeiro, Rio de Janeiro 21941-972, Brazil}
\address[calgary]{Department of Physics and Astronomy, University of Calgary, Calgary AB, T2N 1N4, Canada}
\address[triumf]{TRIUMF, 4004 Wesbrook Mall, Vancouver BC, V6T 2A3, Canada}
\address[ubc]{Department of Physics and Astronomy, University of British Columbia, Vancouver BC, V6T 1Z4, Canada }
\address[tokyo]{Department of Physics, University of Tokyo, Tokyo 113-0033, Japan}
\address[stockholm]{Fysikum, Stockholm University, SE-10609, Stockholm, Sweden}
\address[york]{Department of Physics and Astronomy, York University, Toronto, ON, M3J 1P3, Canada}
\address[liverpool]{Department of Physics, University of Liverpool, Liverpool L69 7ZE, United Kingdom}
\address[auburn]{Department of Physics, Auburn University, Auburn, AL 36849-5311, USA}
\address[negev]{Department of Physics, NRCN-Nuclear Research Center Negev, Beer Sheva, IL-84190, Israel}
\address[riken]{Atomic Physics Laboratory, RIKEN, Saitama 351-0198, Japan}
\address[tokyo2]{Graduate School of Arts and Sciences, University of Tokyo, Tokyo 153-8902, Japan}

\maketitle

\section{Introduction}
Antihydrogen, the bound state of an antiproton and a positron, is the simplest pure antimatter atomic system.
The potential of spectroscopic measurements to probe matter-antimatter equivalence and CPT symmetry has driven a focused experimental effort to study cold antihydrogen atoms.
The first cold antihydrogen atoms were produced by the ATHENA Collaboration \cite{ATHENA_Nature} at the Antiproton Decelerator at CERN, and shortly thereafter by ATRAP \cite{ATRAP_Observation}.
In these and later experiments, the neutral antihydrogen produced was not confined by the Penning-Malmberg traps used to hold the constituent antiprotons and positrons as non-neutral plasmas.
Instead, the antihydrogen atoms either escaped to strike the matter making up the apparatus and annihilate, or were ionised by the electric fields present within the trap volume.
Before precision spectroscopy and other measurements can be carried out, it is highly desirable to first produce a long-lived sample of antihydrogen in an atomic trap.

\section{Apparatus} \label{sec:apparatus}
\begin{figure}
	 \includegraphics[width = \columnwidth]{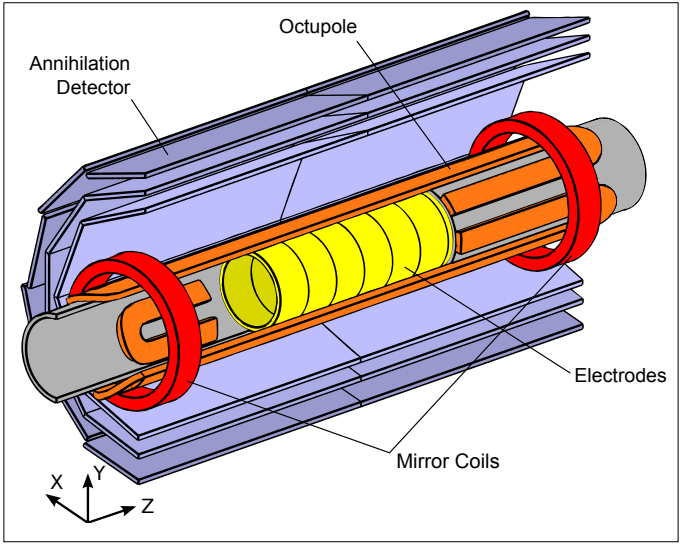}
	\caption{A schematic, cut-away diagram of the antihydrogen production and trapping region of the ALPHA apparatus, showing the relative positions of the Penning-Malmberg electrodes, the minimum-B trap magnets and the annihilation detector. The components are not drawn to scale.}
	\label{fig:apparatus}
\end{figure}

Neutral atoms or anti-atoms can be trapped by exploiting the interaction of their magnetic dipole moments with an inhomogeneous magnetic field.
The ALPHA apparatus (Fig. \ref{fig:apparatus}) produces a magnetic field with a three-dimensional minimum using a variation of the Ioffe\hyp{}Pritchard configuration \cite{Ioffe_Pritchard}.
The quadrupole in the typical Ioffe-Pritchard trap has been replaced by an octupole \cite{ALPHA_MAGNET}.
For the same trap depth, a higher-order multipole produces a smaller transverse magnetic field near the axis of the Penning-Malmberg trap, and has a smaller perturbative effect on stored non-neutral plasmas \cite{Joel_multipole}.
The trap is completed with two short solenoids or `mirror coils' in the longitudinal direction.
The magnets are constructed from niobium-titanium superconductor wound directly onto the wall of the vacuum chamber and are immersed in a bath of liquid helium at 4~K.

The magnetic minimum trap is superimposed on the central region of a Penning-Malmberg trap which confines the antiproton and positron clouds prior to, and during, antihydrogen production.
A uniform solenoidal magnetic field of 1~T ensures radial confinement of the charged particles, while electric fields trap the particles longitudinally.
The electrodes used to produce the electric fields are cooled to approximately 7.5~K by thermally anchoring them to the liquid helium bath used to cool the superconducting magnets.
Compatibility of this combined device with the requirements of storing non-neutral plasmas and producing antihydrogen while the magnetic trap is energised has previously been demonstrated \cite{ALPHA_PlasmasInMutipole}, \cite{ALPHA_HBarInOct}.

The magnetic field strength of the octupole at the inner surface of the Penning-Malmberg trap electrodes is 1.55~T when energised to its operating current of 900~A.
This combines with that of the mirrors (1.2~T at 600~A) and the solenoidal bias field (1~T) to produce a potential well with depth $0.6~\mathrm{K} \times k_\mathrm{B}$ ($k_\mathrm{B}$ is Boltzmann's constant) at the surface of the electrodes for ground state antihydrogen.

The magnets and their external circuitry have been designed to allow for a fast shutdown of the magnetic trap.
This allows any trapped antihydrogen to escape and be detected over a short time interval, thus reducing the background from cosmic rays.
An insulated-gate bipolar transistor (IGBT) acts as a switch to force the current in the magnet to flow in an external resistor network where the energy is dissipated as heat.
This process induces the superconductor to `quench' (make the transition to a normal conductor for a brief period of time).
The current flowing in each magnet is monitored by measuring the voltage drop across a shunt resistor connected in series with the power supply, and, when the fast shutdown is initiated, we measure an exponential decay of the current with time constants of 9.0~ms for the octupole and 8.5~ms for mirror coils.

Antiproton annihilations are identified using a silicon vertex detector \cite{Makoto_detector}.
The charged products of an annihilation, principally pions, can ionise and leave charge deposits in materials they pass through.
The ALPHA detector comprises 60 modules, arranged in three layers in a cylindrical fashion around the mixing and trapping region (see Fig. \ref{fig:apparatus}).
In each module, a double-sided silicon wafer is divided into 256 strips, of widths 0.9~mm and 0.23~mm in the $z$ and $\phi$ directions respectively, oriented in perpendicular directions on the p- and n- sides.
Each strip can be individually addressed to measure the amount of charge deposited.
Charge exceeding a defined threshold causes the electronics controlling that module to output a digital signal, monitored by a control system.

A coincidence of signals from at least two modules in a 400~ns time interval prompts readout and digitisation of the charge collected on all of the detector strips.
Each readout of the detector is referred to as an `event'.
Strips through which particles passed are identified by charge deposits above noise, with a 96\% detection efficiency determined in studies with cosmic rays.
The intersection of two orthogonal hit strips defines a `hit', or the location that an annihilation product passed through the silicon wafer.
Tracks are constructed by fitting a helix to combinations of three hits, where one hit is drawn from each of the layers of detector modules.
Only tracks that produce helices that conform to the expected characteristics of annihilation products are accepted and used to determine the annihilation vertex as the point which minimises the distances of closest approach.
Our system achieves a maximum readout rate of 170~Hz.

A similar detector, made of two layers of silicon and a component sensitive to positron annihilations, was used in ATHENA to identify the first cold antihydrogen atoms \cite{ATHENA_Imaging}.
In ALPHA, the magnet windings are located between the production region and the annihilation detector, and there is a significant chance that a charged particle produced in an annihilation will scatter, reducing the reconstruction performance \cite{Makoto_detector}.
A third layer of silicon, which allows the tracks to be fitted with curves, rather than straight lines, helps to alleviate this, and experimental tests and Monte-Carlo simulations have demonstrated that the detector can reconstruct annihilation vertices with a resolution of better than 1~cm (one sigma).
Space constraints and the low efficiency of detection of gamma photons through the scattering material precluded the addition of a detector for positron annihilations in ALPHA.

\begin{figure}[tbh]
		 \includegraphics[width = \columnwidth]{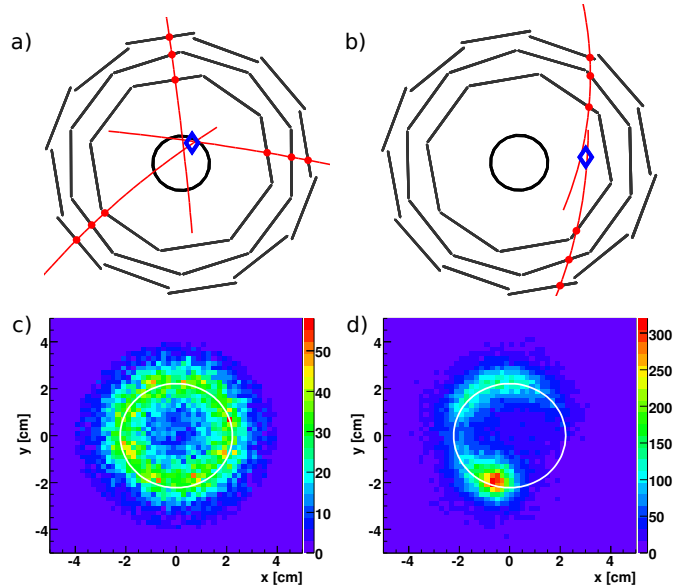}
	\caption{(a) an example reconstruction of an antiproton annihilation and (b) a cosmic ray event. The blue diamond indicates the position of the reconstructed vertex, the red dots the positions of the detected hits, and the inner circle shows the radius of the Penning-Malmberg trap electrodes. (c) The spatial distribution of approximately $2 \times 10^4$ antihydrogen atoms identified in this experiment, projected along the z-axis. The distribution is approximately azimuthally uniform and concentrated around the surface of the electrodes, indicated by the white circle. Small non-uniformities are interpreted to be due to the escape of field-ionised antihydrogen (see text). The escape of bare antiprotons tends to produce highly non-uniform distributions, such as that in (d).}
	\label{fig:reconstruction}
\end{figure}

An example of a reconstructed antiproton annihilation is shown in Fig. \ref{fig:reconstruction}(a).
The detector is also sensitive to charged particles from cosmic rays, which pass though the detector in a straight line and typically reconstruct as a pair of approximately co-linear tracks, as seen in Fig. \ref{fig:reconstruction}(b).

The detector is used as an indicator of antihydrogen production.
Antihydrogen atoms that annihilate on the electrodes will produce an azimuthally uniform distribution of vertices.
When the octupole field is energised, the trajectories of antiprotons redistributed by the process of antihydrogen production and subsequent ionisation can be unstable, giving rise to a component with eightfold symmetry \cite{ALPHA_HBarInOct}.
The distribution of annihilations measured during the antihydrogen production periods in the trapping experiments described in this letter is shown in Fig. \ref{fig:reconstruction}(c).
An azimuthally uniform component dominates, and a small contribution from an eightfold symmetric pattern is also present.
This can be contrasted to the escape of bare antiprotons, which tend to yield very non-uniform distributions \cite{ATHENA_Imaging}.
An example distribution, created by deliberately destabilising an antiproton plasma, is shown in Fig. \ref{fig:reconstruction}(d).
The octupole magnet was not energised for this measurement.

\section{Method}
\label{sec:Method}
The Antiproton Decelerator delivers approximately  $3~\times~10^7$ antiprotons at an energy of 5.3~MeV every $\sim$100~s.
As they enter our apparatus, the particles pass through a degrading foil of $218~\mathrm{\mu m}$ of aluminium, where approximately $10^5$ are scattered to energies lower than 4~keV and are dynamically trapped between two high-voltage electrodes.
This `catching' region of the apparatus is surrounded by a solenoid which increases the longitudinal magnetic field to 3~T during this process.
The antiprotons cool through collisions with electrons in a 0.5~mm radius, pre-loaded plasma, containing $1.5\times10^7$ particles \cite{Pbar_Electron_Cooling}.
The electrons self-cool through the emission of cyclotron radiation.
The resulting two-component plasma, containing $4.5\times10^4$ antiprotons, is then azimuthally compressed by applying a `rotating-wall' electric field \cite{Rot_Wall}, \cite{ALPHA_RW}.

The solenoidal magnetic field is lowered to 1~T before the particles are transferred to the antihydrogen production or `mixing' region, where the electrodes have been designed with large diameter (44.55~mm) and small thickness (1.5~mm) to place the inner surface of the electrodes as close as possible to the octupole windings, thereby maximising the depth of the magnetic minimum trap.
The voltages applied to the electrodes are generated by low-noise amplifiers and are heavily filtered to reduce the level of electronic noise, which may undesirably heat the plasmas.
A series of electric field pulses is then used to separate the electrons from the antiprotons, taking advantage of the much higher velocity of the electrons.

In parallel with this operation, a plasma of positrons is accumulated from a Na-22 radioactive source and cooled using a nitrogen buffer gas in a Surko-type device \cite{Accumulator} before being transferred to the main apparatus.
There, the number of positrons is adjusted to a desired level and the plasma  is compressed using the rotating-wall technique.

The radial density profiles of each of the types of plasmas can be directly measured by destructively extracting the plasma onto a micro-channel plate/phosphor screen assembly \cite{ALPHA_MCP}.
With knowledge of the potentials applied to the electrodes, the full three-dimensional density distribution and electric potential can be calculated by numerically solving the Poisson-Boltzmann equation \cite{Poisson-Boltzmann}.
By changing the parameters of the rotating-wall compression fields and measuring the effect with the plasma imaging device, we can tailor the plasmas to suit an experiment.
With good reproducibility, plasmas with well-defined radius and density can be prepared.
For the experiment described in this letter, the prepared positron plasma is $1~\mathrm{cm}$ long in the $z$ direction, has a radius of $1~\mathrm{mm}$ and has a peak density of $7 \times 10^7~\mathrm{cm}^{-3}$.
The radius of the axially separate antiproton plasma is 0.8~mm, ensuring complete overlap with the positrons.

The temperatures of the plasmas can be measured by slowly (with respect to the axial oscillation frequency) releasing the plasma from the confining well.
The rate of collisions amongst the plasma particles is high enough to ensure that the temperatures in the parallel and perpendicular degrees of freedom have equilibrated.
In thermal equilibrium, the first particles to be released will be drawn from the tail of a Boltzmann energy distribution, which can be approximated by an exponential.
By measuring the number of particles released as a function of well depth, we can thus determine the  temperature from an exponential fit \cite{temperature}.
Analysis of the dynamics of the plasma through the measurement process suggests that the temperature obtained from this method will be higher than the true temperature by a factor between 1.5 and 2 for the positron plasma, and around 15\% for the antiproton plasma.
However, we do not apply the corrections calculated from this analysis to temperatures reported in this letter.

The positron and antiproton clouds are placed in adjacent potential wells in a variation of the nested-trap arrangement \cite{NestedTrap}.
The space-charge ($\sim$2.1~V) of the positron plasma fills most of the central well so that the antiprotons and positrons are separated by a potential barrier of approximately 500~mV.
Before combining the antiprotons and positrons, the magnetic trap is fully energised.

We find that the temperature of a stored electron or positron plasma does not automatically match that of the cryogenic surroundings, as might be expected if all heating sources are ignored.
Once the positrons have been placed in the nested well and the magnetic trap energised, we measure their temperature to be 71~K $\pm$ 10~K.
Once the antiprotons have been separated from the cooling electrons, they are no longer effectively cooled, and we measure their temperature to be 358~K $\pm$ 55~K.
The uncertainty quoted here is one standard deviation of a collection of measurements.

Injection of the antiprotons into the positron plasma is achieved by autoresonantly exciting the motion of the antiprotons parallel to the magnetic field.
The well confining the antiprotons is anharmonic; thus, the oscillation frequency is a decreasing function of the oscillator energy.
Initially, the antiprotons are confined to the bottom of the well, and have an oscillation frequency close to the linear well oscillation frequency.

To inject the antiprotons, we apply a sinusoidal drive whose frequency sweeps downwards through the linear well oscillation frequency.
With appropriate choice of drive parameters, the antiprotons autoresonantly lock to the drive, such that their longitudinal energy adjusts to keep their oscillation frequency matched to the drive frequency \cite{AR}, \cite{Joel_AR}, \cite{Barth_AR}.
This method allows the parallel energy of the antiprotons to be quickly and precisely changed with little impact on the transverse energy.

When the antiprotons have sufficient energy to enter the positron plasma, an abrupt change in the character of the antiproton orbits occurs, with a corresponding abrupt change in the antiprotons' oscillation frequency, and they decouple from the drive and cease to resonantly gain energy from the drive.
We employed a 1~ms long drive, swept from 320~kHz to 200~kHz, producing a $\sim$55~mV amplitude oscillation on the trap axis.
This injects 70\% of the antiprotons in approximately 200~$\mu$s, with the remainder staying below the energy needed to enter the positrons.
For reasons that are not yet well understood, the temperature of the positron plasma increases to 194~K $\pm$ 23~K after injection of the antiprotons.
Collision calculations indicate that the antiprotons quickly equilibrate to this temperature while inside the positron plasma \cite{pbarSlowing}.
We note that the mixing scheme used ensures that the antiprotons have little kinetic energy as they pass through the positron plasma.
This is in contrast to the experiments discussed in reference \cite{ATHENA_spatialDist},  where the antiprotons had kinetic energies $>$ 10~eV parallel to the magnetic field, and which showed evidence that the antihydrogen velocity distribution did not correspond to thermal equilibrium between the antiprotons and the positron plasma.
Carrying out a similar analysis for the antihydrogen distribution measured in our experiment supports the claim that the antiprotons quickly come into equilibrium with the positron plasma.

Once inside the positron plasma, the antiprotons can combine with the positrons to form antihydrogen atoms.
As in \cite{ALPHA_HBarInOct}, most of the antihydrogen has a kinetic energy too high to be trapped and escapes the trap volume to annihilate on the surrounding apparatus or is ionised.
The interaction is allowed to continue for 1~s, during which we observe 2700 $\pm$ 700 annihilation counts.

Following the analysis in \cite{ALPHA_HBarInOct}, we consider two possible sources of annihilations -- antihydrogen striking the wall, or an antiproton that has formed weakly-bound antihydrogen, been ionised by the electric fields at high radius, and guided by the magnetic field to the wall.
Comparing the vertex distribution to that obtained when the octupole is not energised allows us to estimate the fraction of the annihilations that correspond to antihydrogen that is strongly enough bound to not be ionised before reaching the wall.
From this procedure, we estimate that between 70\% and 85\% of the counts are due to antihydrogen strongly enough bound to survive at least one pass through the electric fields.

Following the 1~s mixing period, uncombined antiprotons and positrons are ejected from the trap by manipulating the confining potentials.
A series of electric field pulses is applied across the length of the trap to clear any charged particles.
However, as will be discussed in section \ref{sec:mirrorTrap}, the inhomogeneous magnetic field can cause some charged particles with extreme energies to remain, possibly mimicking trapped antihydrogen.
Electrostatic barriers at either end of the trapping volume prevent these antiprotons escaping along the trap axis, though some can still escape radially.
Removal of the charged particles takes 80~ms, following which the neutral trap remains energised for a further 50~ms before the shut-down of the magnets is triggered.
As the magnetic field falls, any antihydrogen held in the trap will escape and annihilate on the surrounding apparatus.
After 30~ms, the depth of the magnetic minimum has fallen to less than 0.1\% of its initial value; this defines the time window in which we search for escaping trapped antihydrogen.

In a systematic search for trapped antihydrogen, ALPHA conducted this experiment 212 times, in addition to experiments used for diagnostics and controls, over a three-week period in late October and early November 2009.
Before interpreting the results of these experiments, we first describe the method of discriminating between antihydrogen annihilations and the background.

\section{Annihilation Identification and Cosmic Ray Rejection}
\label{sec:cosmics}

Our experiment is designed to identify trapped antihydrogen by releasing it from the magnetic trap and detecting the antiproton annihilation as it strikes the surrounding apparatus.
It is vitally important in this scheme to have a sensitive and efficient method of distinguishing annihilations from cosmic rays.

Antiproton annihilations and cosmic rays exhibit distinct features characterised by their event topology (see Fig. \ref{fig:reconstruction}(a) and (b).)
If our reconstruction algorithm successfully identifies an annihilation vertex, we use the topology of the event to distinguish antiprotons from the cosmic background.
Our analysis procedure parameterises each event in terms of the number of tracks present, the radial position of the vertex and the squared residual from a linear fit to the hit positions, and accepts or rejects it as an annihilation based on the values of these parameters relative to thresholds or `cuts'.

These features were studied by collecting a sample of cosmic ray events, consisting of approximately $3.1 \times 10^5$ events, recorded over 20 hours during which no antiprotons were delivered to the apparatus (an average rate of 4.31~Hz).
We compared this sample to approximately $2.4 \times 10^4$ events recorded during a total of 170~s of antihydrogen production in the magnetic trap.
These events overwhelmingly consist of antihydrogen annihilations on the surface of the Penning trap electrodes.

These samples were used to choose and optimise the method for identifying antiproton annihilations, without making reference to the data recorded during the trapping experiments, so the possibility of experimenter bias influencing the analysis was eliminated.
Use of real, measured distributions to develop the selection criteria is superior to a model-based approach -- Monte-Carlo simulation -- since many of the systematic effects are automatically taken into account.

\begin{figure}
	 \includegraphics[width = \columnwidth]{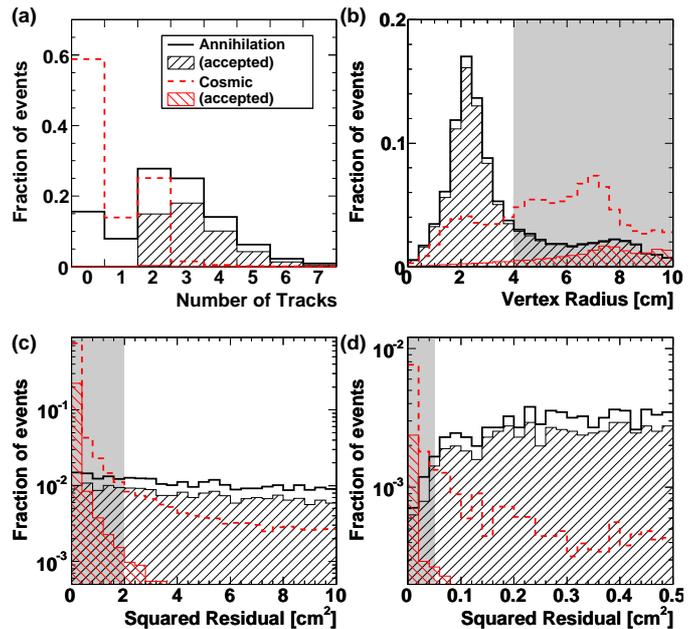}
	\caption{The measured distributions of (a) the number of identified charged particle tracks, (b) the radial coordinate of the vertex, and the squared residual from a linear fit to the identified positions for the events with (c) two tracks and (d) more than two tracks. The inset in (a) shows the region close to the y=0 axis. Each distribution has been normalised to the total number of events in the calibration sample. The distributions from annihilations are shown in solid black lines and from cosmic rays in dotted red lines. The hatched regions below each line are the distributions remaining after the cuts are applied. In (a), the accepted cosmic-ray distribution is too small to be distinguished from the x-axis. The solid shaded regions indicate the range of parameters that are rejected to minimise the p-value (see text). In these regions, the cut for the parameter for which the dependence is shown is ignored.}
	\label{fig:detectorCuts}
\end{figure}

Fig. \ref{fig:detectorCuts}(a) compares the distributions for the number of tracks observed in an event for each of the two samples.
95\% of the cosmic events have two or fewer tracks, while 58\% of annihilations have at least three.
We interpret the cosmic events with more than two tracks to be due to events in which a spurious track has been identified, or to events in which the cosmic particle produces a shower of particles also following the direction defined by the initial momentum.
Events in which no tracks are identified include events in which read out was triggered by electronic noise in the control system.

A significant fraction of antiproton annihilations can have two tracks, so it is not desirable to reject all such events as background.
Instead, we make use of the fact that the high-momentum cosmic rays will tend to be deflected only slightly as they pass through the apparatus and the magnetic field.
The detected hits from a cosmic ray will tend to lie on a straight line, while those from the products of an annihilation will, generally speaking, not.

Such cosmic rays can be identified by assessing how well any two tracks are compatible with one, straight-line track.
We find the straight line that best fits the hit positions making up two of the particle tracks, and calculate the sum of the squared residual distances from each hit to this line.
The distributions of this quantity for each of the  samples is shown in Fig. \ref{fig:detectorCuts} where the events have been separated into those with two tracks (\ref{fig:detectorCuts}(c)), and those with at least three (\ref{fig:detectorCuts}(d)).
We see that almost all of the events from the cosmic sample have small values of the squared residual, while the distribution from antihydrogen annihilation is more featureless, allowing cosmic rays to be rejected by requiring a large squared residual value. 
Large values of the squared residual correspond to curved tracks or tracks at an angle to each other, which are seen in annihilation events, but not in cosmic rays.

Antihydrogen annihilations are expected to occur on the inner surface of the Penning-Malmberg trap electrodes, which is the first solid matter material an atom of antihydrogen will encounter as it escapes the magnetic trap.
Cosmic rays, on the other hand, should pass through the apparatus at a random radius.
Fig. \ref{fig:detectorCuts}(b) shows the distribution of the reconstructed annihilation vertices as a function of distance from the trap axis (the radius in cylindrical coordinates).
The distribution from antihydrogen annihilations is peaked at approximately 2.2~cm, which corresponds well to the radius of the electrodes.
Events with vertices far away from this region are more likely to be cosmic rays.

Optimisation of the placement of the cuts on the event parameters was performed by minimising the expected p-value by varying the cut thresholds and examining the effect on the p-value, assuming a signal rate based on a preliminary survey of the data.
(The p-value is the probability that statistical fluctuation in an expected background gives rise to the observed result in data \cite{pdg}.)
This was achieved by requiring that the position of the annihilation vertex lie within 4~cm of the trap axis, as well as the requirement that for events with at least three tracks, the squared residual should exceed 0.05~$\mathrm{cm}^2$, while for events with fewer tracks, the squared residual to a straight-line fit should exceed 2~$\mathrm{cm}^2$.
After applying the selection criteria, the distributions of the event characteristics change -- the distributions for the accepted events are also shown in Fig. \ref{fig:detectorCuts}.
The majority of events from the cosmic sample that remain after applying this process are two-track events that appear to have undergone scattering in the material of the experiment so that they have a large squared residual value.

The overall efficiency of detecting an annihilation is estimated to be $(42 \pm 7)\%$ from the product of trigger efficiency $(85 \pm 15)\%$, the fraction of events that produce a vertex $(74.6 \pm 0.5)\%$, and the acceptance of the final cuts $(65.7 \pm 0.6)\%$.
The uncertainty on the trigger efficiency is almost entirely a normalisation uncertainty in the scintillation detector used to measure the absolute number of antiprotons in the trap, while the other uncertainties are statistical.
The cuts accept $(0.51 \pm 0.01)\%$  of the cosmic background, corresponding to  an absolute rate of $(2.2 \pm 0.1) \times 10^{-2}~\mathrm{Hz}$, as determined from the cosmic sample.
The total observation time for the 212 trapping experiments was 6.36~s, implying an expected cosmic background of $0.14 \pm 0.01$ events passing the selection criteria.

\begin{figure}[tb!]
	\includegraphics[width = \columnwidth]{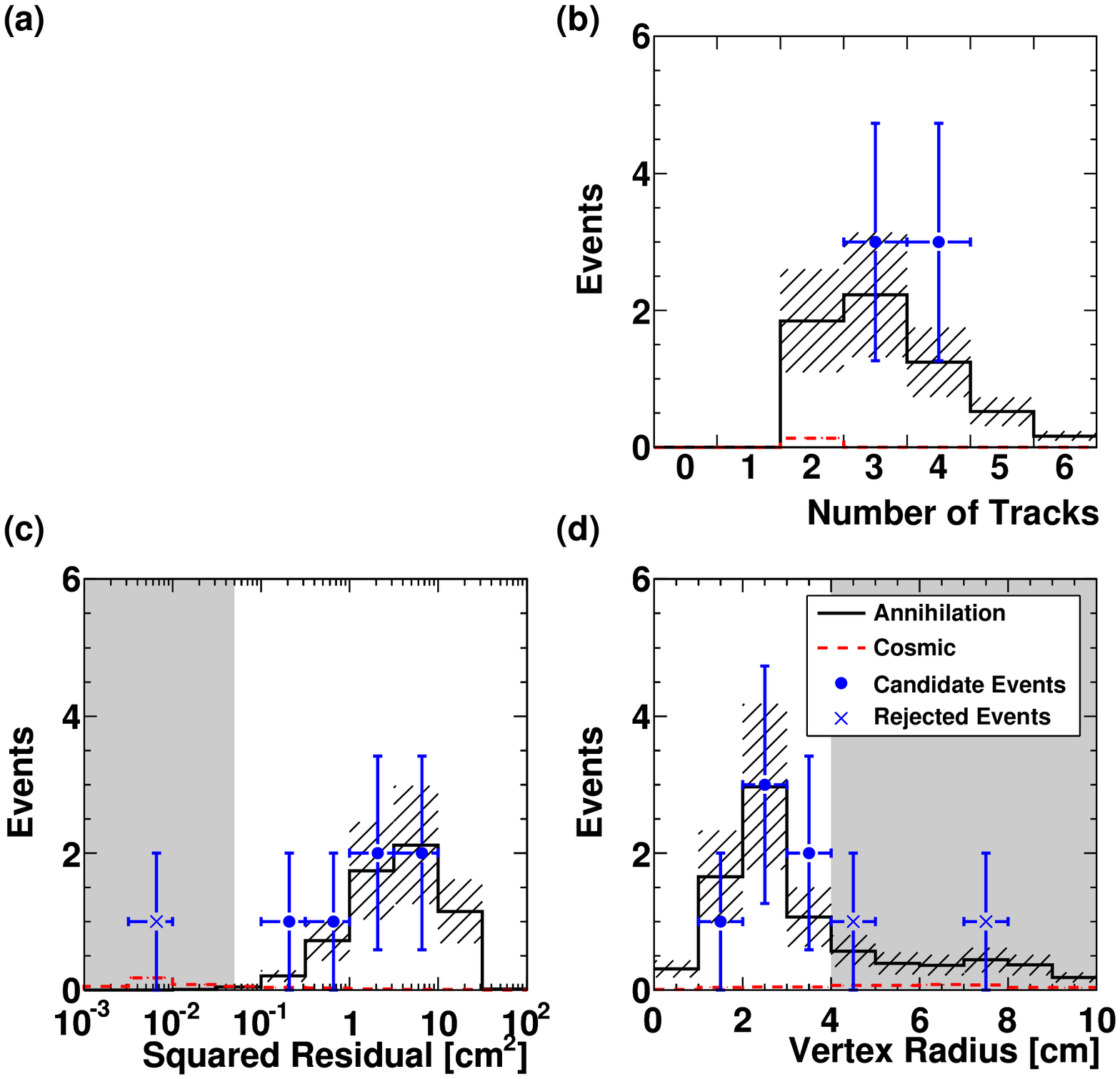}
	\setlength{\unitlength}{\columnwidth}
	\put(-0.95, 0.55){\includegraphics[width = 0.4\columnwidth, clip, trim = 5cm 5cm 5cm 5cm]{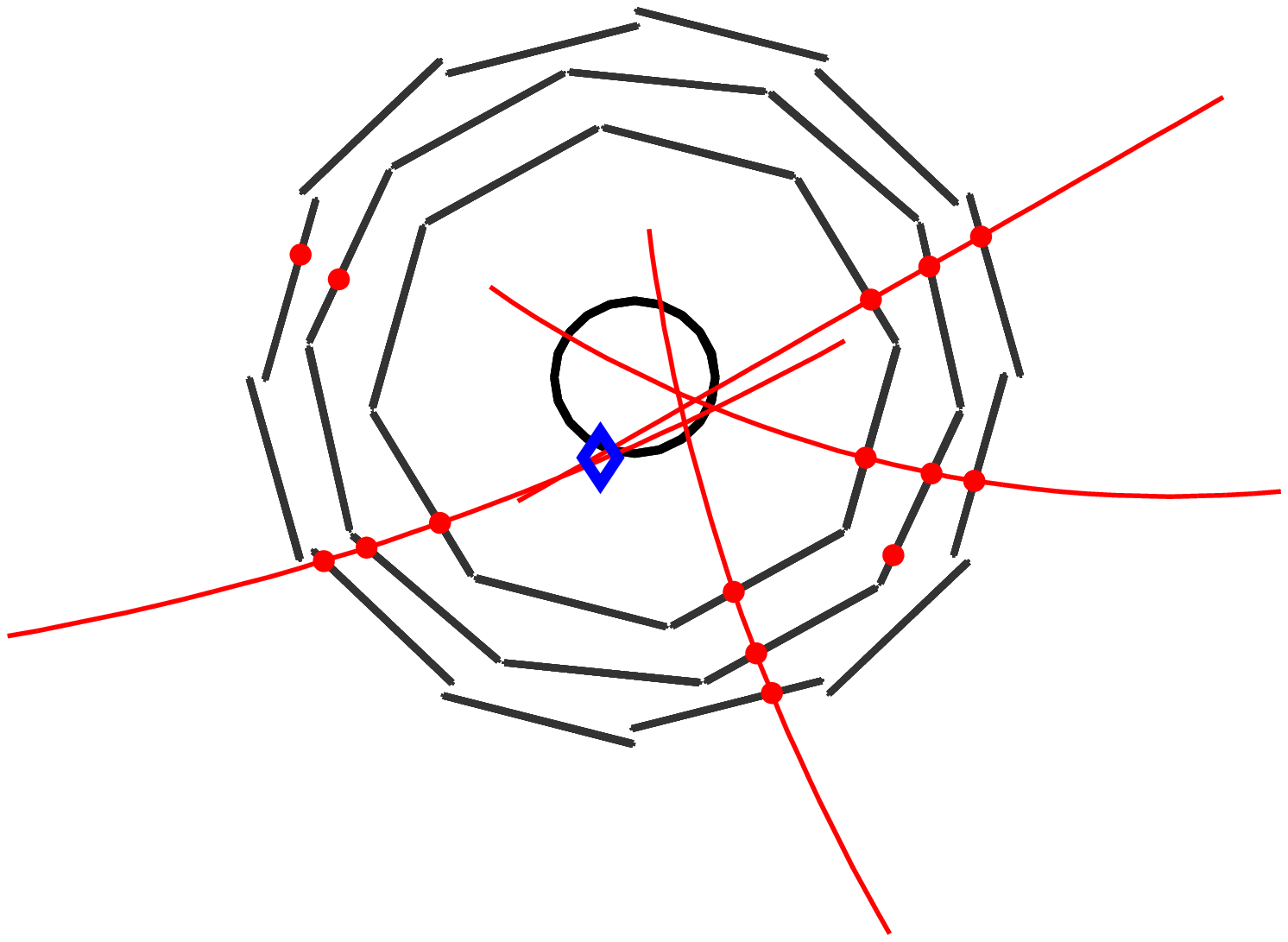}}
	\caption{(a) A view of the reconstruction along the $z$-axis for one of the six events that passed the selection criteria. The elements of this figure have been described in Fig. \ref{fig:reconstruction}. Comparisons of the events measured in the trapping experiment with the distributions of (b) number of tracks per event (c) squared residual and (d) vertex radial position obtained from applying the selection criteria to the calibration samples. The cut on squared residual and radial position has been ignored in the grey shaded regions excluded by the selection criteria in (c) and (d) respectively. The distributions are obtained by scaling the calibration sample distributions so that the expected number of events, given our data -- six for the annihilation data and 0.14 events for the cosmic events -- fall inside the acceptance region. The hatch-marked uncertainty regions represent the uncertainties on these numbers. The error bars for the measured events are the counting errors. }
	\label{fig:cuts_v_Data}
\end{figure}

Throughout the trapping series, 36 events were recorded in the 30~ms time window, and when the cuts were applied to this data, six survived the selection criteria.
The probability of this observation being due to fluctuations in the cosmic background (the p-value) is $9.2 \times 10^{-9}$, corresponding to a significance of 5.6 standard deviations.
A view of the reconstruction of one of these events is shown in Fig. \ref{fig:cuts_v_Data}(a).

\begin{figure}[tb!]
	\includegraphics[width = \columnwidth]{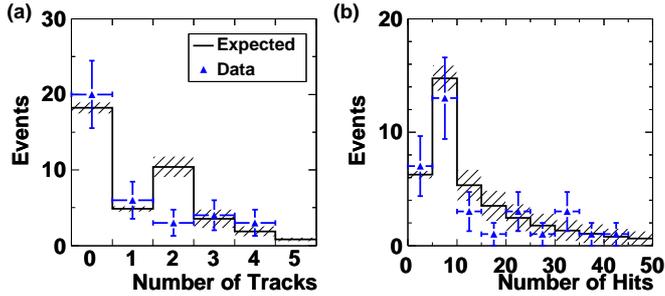}
	\caption{Comparisons of the events measured in the trapping experiment with the distributions of (a) number of tracks per event and (b) number of hits per event, produced by combining the calibration samples with the expected weights (see text). The uncertainties in the distributions are based on the uncertainties in the weighting factors, and the error bars on the data are the counting errors. }
	\label{fig:cuts_v_Data2}
\end{figure}

To ensure that the calibration samples are representative of the data collected during the trapping experiments, we carried out careful comparisons of the detector performance during collection of these data samples to the performance during the trapping experiments, as well as to background windows immediately before and after the observation window.
We did not observe any significant difference in characteristics such as the number of tracks and hits per event, or the level and spread of the voltages read from the silicon module strips.
We also compared antihydrogen annihilation samples at different annihilation rates to ensure that the characteristics do not depend on the rate at which the detector is triggered or read out.
All of our analyses and cross-checks show consistency between the calibration samples and the data collected during the trapping experiments, and we conclude that it is valid to use these samples to adjust the selection criteria and as an estimate of the background.
Comparisons of the cosmic sample and background windows in the trapping experiment data limit the size of any systematic effects to a negligible level. 

In Fig. \ref{fig:cuts_v_Data} (b)--(d), we show the distributions of event parameters from the calibration samples, scaled so that the expected number of events based on our measurements fall into the acceptance region.
For the annihilation sample, six events were accepted, while for the cosmic sample, the measured cosmic rate and the acceptance of the selection criteria imply that 0.14 cosmic events would have been accepted.
The selection criteria have been applied to the trapping experiment events and calibration distributions, except in the shaded grey rejection regions in (c) and (d), where the residual and radial thresholds, respectively, have been ignored.
The surviving events are shown as solid circles, while the events that are rejected by the residual and radial thresholds in (c) and (d) respectively are shown as crosses.

Fig. \ref{fig:cuts_v_Data2} shows similar comparisons to Fig. \ref{fig:cuts_v_Data}, but for all of the events recorded in the trapping experiments (i.e. not applying the selection criteria).
The parameters plotted are (a) the number of tracks and (b) the number of hits per event.
The expected distribution is calculated by combining the cosmic distribution from the calibration samples, scaled by the rate of events in the cosmic sample and the length of the observation window, for a total of 27.4 events, and the annihilation distribution, scaled by the number of events that pass the selection criteria, and corrected for the acceptance efficiency, giving $12.2 \pm 5.0$ events.

The plots shown in Fig. \ref{fig:cuts_v_Data} and Fig. \ref{fig:cuts_v_Data2} demonstrate that the distributions measured in the trapping experiment are consistent with the calibration samples, within the statistics.
There appears to be a deficit in the number of events with exactly two tracks, but not to a point incompatible with statistical variation -- approximately 11\% of sets of six annihilation events would be expected to have no events with two tracks.
The overall agreement between the expected and measured distributions support the validity of our analysis.

Each of the experimental runs in which an event survived the cuts was closely examined, and the set of six was verified to be representative of the complete 212.
No anomalies were found in, for instance, the readings from environment monitoring sensors, positron or antiproton source performance, the background rates of the vertex detector, or the number of annihilations recorded during antihydrogen production.

Concurrent with the trapping experiment, a number of control experiments were carried out.
For instance, 121 repetitions of the experiment were carried out without antiprotons in the trap, verifying that the transient electromagnetic fields caused by the quench of the magnets do not induce false annihilation signals.
A further 40 were carried out with only positrons present, to ensure that positron annihilations cannot mimic detection of an antiproton.
No events meeting the criteria for selection as an annihilation were identified in these runs.
If the same process that caused us to identify annihilation events in our trapping experiment was present at the same rate, there would be only a $\sim 1\%$ chance of not observing any events in this data.
This is an additional strong indication that our signal is not due to cosmic rays.

The conclusion of this analysis is that we have, to a high level of certainty, observed antiproton annihilations on release of our magnetic trap.
In the next section, we will address the possible sources of antiproton annihilations that are not due to trapped antihydrogen atoms.

\section{Mirror trapping of bare antiprotons} \label{sec:mirrorTrap}

The observation of an annihilation does not immediately imply the presence of antihydrogen.
The magnetic minimum trap can also act as a trap for charged particles, including bare antiprotons.
This arises from the adiabatic conservation of the magnetic moment of a gyrating particle, $\mu = E_\perp/B,$ where $E_\perp$ is the kinetic energy of the particle in the plane transverse to the local magnetic field and $B = |\mathbf{B}|$ is the magnitude of the magnetic field.
In a magnetic field that varies along the trajectory of a particle, this yields an equation for the parallel speed
\begin{equation}
	v_\parallel^2 = v_0^2 \left(1 -\frac{v_{\perp,0}^2}{v_0^2} \; \frac{B}{B_0} \right) ,
\end{equation}
where $v_0 = \sqrt{v_{\perp,0}^2+v_{\parallel,0}^2}$ is the speed of the particle at a point where the magnetic field is $B_0$.
It can be seen that for sufficiently high $B/B_0$ or $v_{\perp,0}/v_0$, $v_\parallel$ will reach zero, which corresponds to a turning point in the motion, and so the particle is `reflected' from a region of increasing magnetic field.
This effect is called `mirror-trapping'.

After removing the bulk of the antiprotons by manipulating the potentials, a further series of pulses is applied across the mixing region, before the shutdown of the magnetic trap, to remove any that are mirror-trapped.
Each pulse has an average electric field of approximately $2.5~\mathrm{V\:cm^{-1}}$ and is applied for 10~ms.
We observe annihilations corresponding to a few tens of antiprotons coincident with the pulses per experiment, showing that antiprotons can indeed become mirror-trapped, and that this method can remove at least some of them.

The effect of the pulses can be considered by combining the electrostatic potential energy, $(-e)\Phi$, with the potential energy from the interaction of the particle's magnetic moment with the magnetic field to give a pseudo-potential of the form
\begin{equation}
\label{eq:pseudoPotential}
	U = E_{\perp,0} \; \left(\frac{B-B_0}{B_0}\right) + \left(-e\right) \Phi,
\end{equation}
where the constancy of $\mu$ has been exploited to rewrite $E_\perp$ in terms of the minimum transverse kinetic energy, $E_{\perp,0}$ and the minimum magnetic field $B_0$.

\begin{figure}
	 \includegraphics[width=\columnwidth]{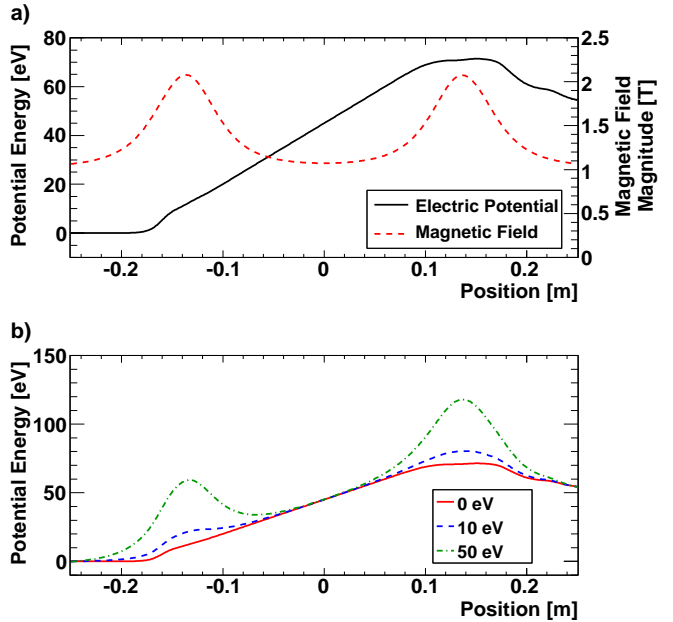}
	\caption{The electric potential (solid line in (a)) and the interaction of the magnetic moment of the antiproton with the inhomogeneous magnetic field (dashed line in (a)) combine to give an $E_\perp$-dependent pseudo-potential on the axis of the trap, three examples of which are shown in (b). Particles with low transverse kinetic energy are not significantly affected by the magnetic fields and are easily cleared from the mixing region. Particles with higher $E_\perp$ are the more strongly influenced by the magnetic field and can be confined.}
	\label{fig:pseudoPotential}
\end{figure}

The on-axis magnetic field of the magnetic minimum trap and electric potential generated by the `clearing' pulses, and the pseudo-potentials for a range of particle transverse kinetic energies are shown in Fig. \ref{fig:pseudoPotential}.
Low $E_\perp$ particles are ejected from the trap by the electric field which overcomes the force due to the inhomogeneous magnetic field.
At higher values of  $E_\perp$, a local minimum develops in the pseudo-potential, where particles can remain trapped.
By following the time evolution of this potential through the particle extraction procedure, we estimate that a particle must have $E_\perp$ of at least 20~eV to be trapped.

Off the axis of cylindrical symmetry, the influence of the octupolar magnetic field produces complex trajectories which cannot be subjected to the same analysis.
Instead, the trajectories are calculated numerically.

In our code, a particle is given an initial position and velocity from a pre-selected distribution.
The initial distribution of antiprotons in our experiment (particularly those created by ionisation of antihydrogen) is not well known, so conservative options that enhance the proportion of antiprotons with high $E_\perp$ and high initial radius were chosen.
Spatially, antiprotons are evenly distributed throughout the trap volume.
Two kinetic energy distributions were used -- a three-dimensional Maxwell-Boltzmann distribution with a temperature of 10~eV and a uniform distribution with a maximum of 40~eV.
Using these distributions we estimate that we have simulated of order 100 times more antiprotons with $E_\perp > 10~\mathrm{eV}$ than existed in our 212 trapping experiments.

To ensure consistency and minimise the probability of error, the antiprotons were propagated using two sets of equations of motion - one employing the full 3D Lorentz force and the other using a guiding centre approach, which neglects the cyclotron motion by calculating the motion of the particle's gyrocentre \cite{guidingCentre}.
Each calculation was independently implemented twice, with different integration routines, and all were seen to conserve invariant quantities.
The results of the calculations are in good agreement, but for simplicity, only data generated by the full 3D Lorentz calculation will be shown.

The trajectory of a particle was followed through the simulation until it either crossed the inner wall of the electrodes or survived until at least 4~ms after the end of the electric pulse sequence.
The initial and final positions and velocities were then recorded for further study.

The magnetic field was included by fitting an analytic model to a calculation of the magnetic field using the TOSCA/OPERA3D package \cite{opera} and the real magnet structure.
This model preserves all of the features of the magnetic field and differs from the calculation by at most 0.02~T.

The electric potential produced by the excitations applied to the Penning-Malmberg trap electrodes is calculated using a finite-difference method.
The time varying voltage applied by the electronics chain was calculated from knowledge of the amplifier and circuit design and agrees well with measurements of the voltage at the external vacuum feed-throughs.

Once the bulk of the particles has been removed, the density of particles in the experiment is very low, with an upper limit of approximately $10~\mathrm{cm^{-3}}$, implying a collision rate of less than $10^{-7}$~Hz or $10^{-4}$ collisions per experiment.
We thus neglect inter-particle effects and only simulate single particle trajectories.

The majority of particles simulated were not stably trapped or were removed by the electric field pulses.
We find that no particles with $E_\perp$ less than 20~eV remain in the trap, which is the same conclusion as that reached by our on-axis pseudo-potential analysis, but now applies to the entire trap volume.

We must therefore consider mechanisms that are capable of producing antiprotons with $E_\perp$ of at least 20~eV.
Taking the measured temperature (358~K, 31~meV) and the number of particles in the antiproton plasma, we can calculate the fraction of a thermal distribution with high $E_\perp$ and find that it is several orders of magnitude too small to account for the six observed events.
Thus, only non-thermal sources of antiprotons are important.

Antiprotons with high parallel kinetic energy are relatively easy to produce, as an antihydrogen atom can be ionised and accelerated by the strong electric fields at the edge of the antihydrogen formation region.
In order to convert this parallel kinetic energy into perpendicular energy, the antiproton must undergo a collision.
As has already been described, the rate of antiproton-antiproton collisions is low enough to be neglected.

An antiproton can also undergo a collision with a residual gas atom in the trap.
In the cryogenic environment of the trap, the residual gases are predominantly hydrogen and helium, and the individual atoms have small velocities compared to mirror-trapped antiprotons.
For a `hard' collision, in which approximately $20~\mathrm{eV}$ of energy is transferred to the perpendicular degree of freedom, an incident antiproton must have parallel energy greater than $\sim 30~\mathrm{eV}$.
The density of gas atoms is known from the rate of annihilations of stored particles and the rate of hard collisions with antiprotons can be evaluated numerically to be $\sim 10^{-5}~\mathrm{Hz}$.
We estimate the probability of such an encounter to be at least five orders of magnitude too low to account for six annihilations.

Our estimate of the probability of producing a mirror-trapped antiproton is extremely small.
However, we lack complete knowledge of the spatial and energy distributions of antiprotons during and after the mixing procedure, which means that we cannot rely on these calculations to completely exclude the presence of mirror-trapped antiprotons.

To experimentally test for possible mirror-trapped antiprotons, we carried out a series of measurements that performed the same manipulations on the particles as the main experiment, except that the octupole or one of the mirror coils was not energised.
This is not entirely a valid null experiment, since changing the magnetic field may influence the initial conditions of the particles, the antihydrogen formation process, or the probability of mirror-trapping an antiproton.
No annihilation-like events were observed in these experiments, though the number of experiments performed was small, so the statistical significance is low.
If the same process that caused us to identify annihilations in the trapping experiments was present at the same rate, there would be a probability of 0.17 to observe zero events from statistical fluctuations alone.

Because of the presence of mirror-trapped antiprotons as a background, many of the obvious experiments which can be proposed are not good null tests.
We have not yet identified a null experiment that is feasible at our low rate of events that can influence the rate of trapped antihydrogen without also influencing the background from magnetically trapped antiprotons.

\section{Release signatures of antihydrogen and antiprotons}

Even if mirror-trapped antiprotons are present as we shut down the magnetic trap, this does not necessarily imply that we cannot distinguish them from antihydrogen.
The annihilation vertex detector allows us to determine both the spatial and temporal distributions of annihilations.
We can compare these to what would be expected from the release of antihydrogen and mirror-trapped antiprotons to attempt to draw a distinction between the two.

A ground state antihydrogen atom can have a kinetic energy of no more than 0.6~$\mathrm{K} \times k_\mathrm{B}$ and remain trapped.
Thus, they move slowly (at most 100~$\mathrm{m\:s^{-1}}$), and, while the magnetic field is falling, will transit the trap only a small number of times before they escape.
During each transit, the magnetic field will change significantly and antihydrogen atoms will not have time to explore the entire trap and find the locations where the trap depth is lowest.
Thus, we expect the z-distribution of annihilations of antihydrogen atoms to be relatively broad.

In contrast, mirror-trapped antiprotons have much higher energy, and move faster than the antihydrogen atoms.
They do have time to explore the trap boundaries, and find the location where they are minimally bound -- midway between the mirror coils -- which is where we expect them to annihilate.

To properly simulate the trajectories during the shutdown of the magnetic trap, we must consider the effect of induced eddy currents on the magnetic field decay.
This effect can be well modelled by passing the measured coil currents through low-pass filters, effectively slowing the decay by 15\%.
For antiprotons, we must also include the forces due to the induced electric fields.

The initial parameters for antiprotons are randomly selected from the set of mirror-trapped antiprotons found in section \ref{sec:mirrorTrap}.
As in the real experiment, antiprotons are prevented from escaping along the $z$-axis by electric potentials, and approximately 50\% of the antiprotons remain trapped after the currents in the trapping magnets have decayed to zero.

In addition to the antiproton codes described above, we also developed two independent programs which modelled the trajectories of the antihydrogen atoms subject to the magnetic moment force equation.
The simulations were initialised with antihydrogen atoms evenly distributed in the trap.
These atoms were initially either in the ground state, with maximum kinetic energy of 0.1~meV, or in the $n=25, l=24, m=24$ excited state, with maximum kinetic energy 1.0~meV.
The typical kinetic energy is larger than the depth of the neutral trap, ensuring that all trappable atoms are considered.
Excited states of antihydrogen were allowed to de-excite through spontaneous emission.
The results are not sensitive to the initial atomic state, except that the initially excited distribution gives a higher fraction of trapped antihydrogen.
The atoms are propagated for between 90~ms and 100~ms before the simulated decay of the magnetic field commences, allowing for the exclusion of transiently-trapped atoms and for randomisation of the phase space.
After the quench, the simulation continues until the antihydrogen atoms hit the inner surface of the electrodes, leave the trap through the ends, or a further 50~ms has elapsed.

\begin{figure}
	 \includegraphics[width=\columnwidth]{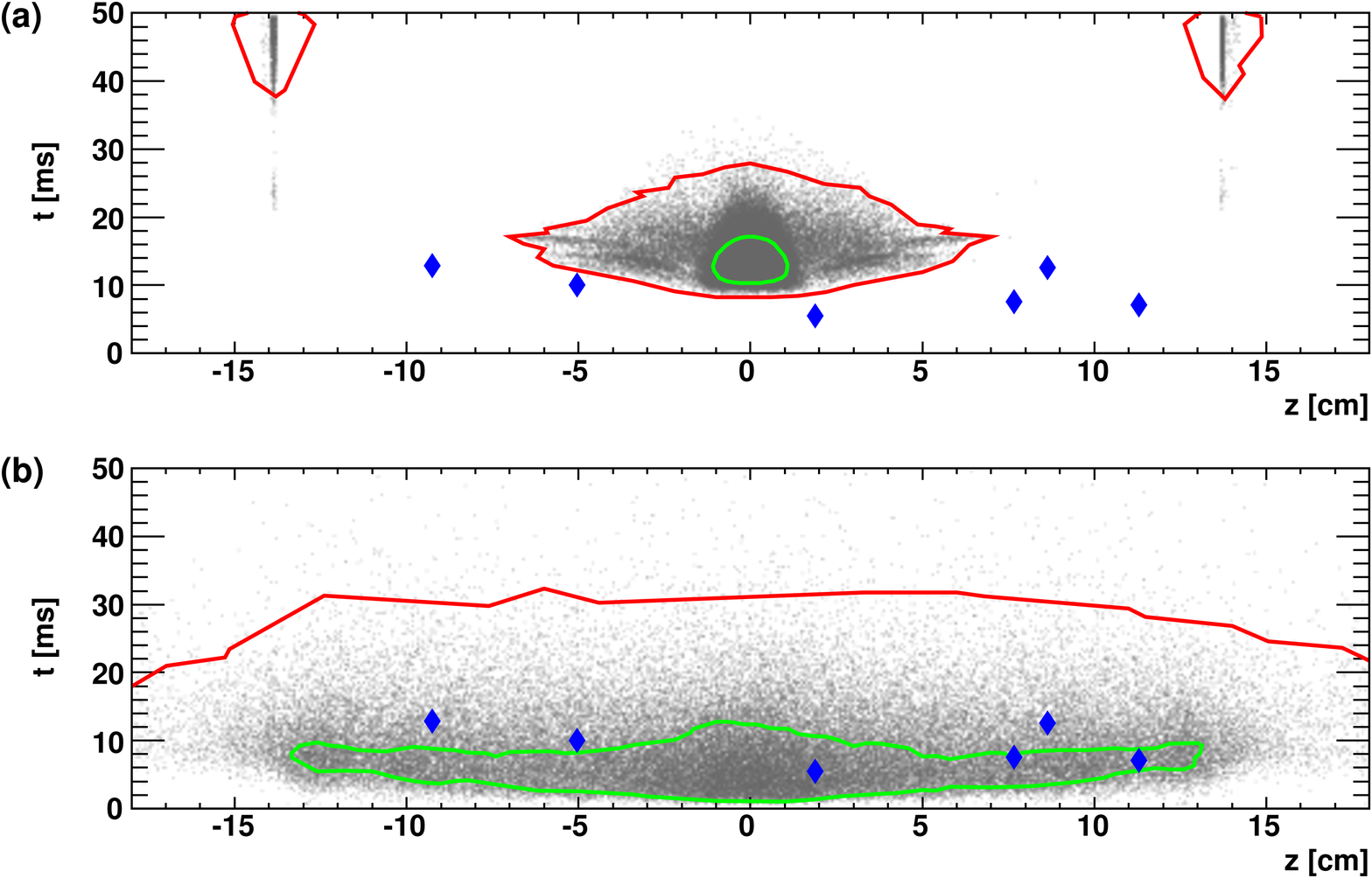}
	\caption{The time after the start of the magnet shutdown and $z$-position relative to the centre of the trap of the simulated annihilations of (a) mirror-trapped antiprotons and (b) antihydrogen atoms released from the magnetic trap. Individual simulated annihilations are shown as discrete points, of which there are 86914 in (a) and 62438 in (b). The lines show the contours of constant density which contain 50\% (green) and 99\% (red) of the density when convolved with the resolution of the detector. The solid diamond-shaped points mark the positions of the six surviving events in the trapping experiment. Some mirror-trapped antiprotons impact at $\pm 14$~cm, where a step in the radius of the electrodes occurs. }
	\label{fig:hitSims}
\end{figure}

The results of these simulations are shown in Fig. \ref{fig:hitSims}.
The horizontal axes show the $z$-coordinate of the position where the particle impacted the electrode wall, and the vertical axes the time of the impact relative to the beginning of the current decay.
When these points are convolved with a function describing the resolution of the annihilation detector, we obtain a continuous density function.
We show the contours of constant density which contain 50\% (green) and 99\% (red) of the integral of this function.
In the transverse plane, the azimuthal resolution of the detector is poor (RMS $\sim 17^\circ$) and most of the useful information, which occurs with a periodicity of $45^\circ$ due to the eightfold azimuthal symmetry  of the octupole's magnetic field, is obscured.
Thus, we do not use the transverse information to compare the simulations with the experiment.

As we expect from the argument above, the distribution of mirror-trapped antiproton annihilations is concentrated near the centre of the trap, in contrast to the much broader distribution of antihydrogen annihilations.

The surviving events clearly lie outside the population of simulated bare antiproton annihilations.
Because the observed points fall far from any simulated annihilation, it is not possible to reliably evaluate a finite value for the probability of observing these events given this data set.
However, all of the surviving events lie outside the 99\% contour; thus, an upper limit on the fraction of sets of six annihilations drawn from the simulated distributions at ($z,t$) positions more extreme than this is $\left(10^{-2}\right)^6 = 10^{-12}$.
On the other hand, we observed that half of the events lie inside the 50\% contour of the antihydrogen distribution, which is the most likely outcome (with a probability of 0.31), indicating that the surviving events are compatible with the release of antihydrogen.

We have varied many parameters in the simulations of mirror trapped antiprotons to observe their effect on the annihilation distribution.
These include introducing tilts and offsets of the magnetic field and using extremely high energy (keV range) antiprotons.
We have not found a distribution for which the observed events are not extremely unlikely.

The results from the simulations are a further strong indication that the surviving events are trapped antihydrogen atoms.
However, without knowing if the simulations carried out fully represent the real experiment, and without an unambiguous control experiment, we cannot yet definitively claim to have observed trapped antihydrogen.

\section{Comparison with theoretical estimates}

Because ALPHA is equipped with the necessary diagnostic devices, the initial experimental conditions are well-determined and reproducible enough to allow us to calculate the expected yield of trapped antihydrogen from this experiment.
Still, there are many effects that must be included, so the final prediction can only be taken as an order-of-magnitude estimate.
We estimate the number of trapped antihydrogen atoms from the simple relationship
\begin{eqnarray*}
N_{\mathrm{detected}} & = & N_{\mathrm{trapped}} \times f_{\mathrm{detection}}\\ & = & N_{\mathrm{produced}} \times f_{\mathrm{0.6K}} \times f_{\mathrm{LFS}} \times f_\mathrm{detection},
\end{eqnarray*}
 where $f_\mathrm{detection}$ is the efficiency to detect a trapped antihydrogen (see section \ref{sec:cosmics}), $f_\mathrm{0.6K}$ is the fraction of atoms produced with energy lower than $0.6~\mathrm{K}\times k_{\mathrm{B}}$, and $f_\mathrm{LFS}$ is the fraction of atoms in trappable `low-field seeking' states.
$N_\mathrm{detected}$, $N_\mathrm{trapped}$, and $N_\mathrm{produced}$ are the number of antihydrogen atoms detected, trapped and produced respectively.
We will discuss each of the terms in turn.

The total number of antihydrogen atoms produced can be estimated by summing the number of annihilation counts over the experimental series and subtracting our estimate (discussed previously in section \ref{sec:Method}) for the number of counts due to antiprotons from atoms ionised by the electric field.
From this, we determine that approximately $4 \times 10^5$ atoms were produced in states low enough to survive the electric fields.

Because the mass of the antiproton is so much larger than that of the positron, the kinetic energy of a newly-formed antihydrogen atom will be close to that of the antiproton before combination, and has two components - the thermal energy and an energy associated with a rotation about the axis of the trap due to the crossed electric and magnetic fields (often called an $\mathbf{E} \times \mathbf{B}$ rotation).
We assume that formation of antihydrogen occurs after the antiproton temperature has equilibrated to the temperature of the positron plasma (194~K).
This is justified since the calculated slowing rate of antiprotons in a positron plasma at this temperature and density \cite{pbarSlowing} is much greater than the antihydrogen production rate, and is consistent with the shape of the spatial antihydrogen annihilation distribution (discussed in section 3).

The $\mathbf{E} \times \mathbf{B}$ rotation rate can be calculated from the solution to the Poisson-Boltzmann equation.
The velocity of the rotation increases further from the axis of the trap, so the radial distribution of points at which antiprotons form antihydrogen can be important.
However, at 194~K the thermal velocity of the antiprotons is the dominant contribution, and we take a simple uniform distribution, for which we numerically evaluate the fraction of antiprotons with kinetic energy less than $0.6~\mathrm{K} \times \mathrm{k_B}$ to be $\sim 1.3 \times 10^{-4}$.

At a temperature of 194~K, the radius of the cyclotron motion of the positron ($\sim 10^{-7}$~m) is larger than the radius of an antihydrogen atom ($\sim 10^{-10}$~m -- $10^{-8}$~m ).
Thus, the distribution of magnetic moments will be purely statistical, and the atoms will be evenly divided between high- and low-field seeking states \cite{Francis_states}.

Using these pieces of information, we can calculate the number of low-field seeking atoms with kinetic energy less than the trap depth of 0.6~K.
This is a simplified, conservative model, in which we neglect some favourable aspects of the cascade from highly excited states.
There is also the possibility that during the cascade, the atom can transition to an unbound state and be ejected from the trap.
However, it has been calculated that this is not probable \cite{Francis_topicalReview}, and we ignore it.

In this simple model, the number of trapped antihydrogen atoms is $\sim 26$.
Scaling by the detection efficiency, we predict that we would identify 11 atoms, which agrees well with our observation of six events.

This model can be extended to include the effects of the cascade from highly excited Rydberg states.
Higher quantum states can have a larger magnetic moment, and the trapping potential will be deeper for these atoms.
As the atom radiatively decays to less excited states (with a lifetime much shorter than the time between antihydrogen production and the observation window \cite{radiativeDecay}), the well depth will reduce, becoming 0.6~K when the atom reaches the ground state.
As described in \cite{cascadeCooling}, antihydrogen atoms that decay near the turning points of their motion will experience a reduction in their total energy.
This results in an effective trap depth of as much as a factor of four higher.
We must also account for the possibility that an atom will reach the ground state without passing through an untrapped state.
A full discussion of the cascade can be found in \cite{Francis_topicalReview} and the references therein.

Including the cascade, we estimate the number of trappable atoms to be $\sim 60$, of which we could identify $\sim 25$.
We see that the inclusion of this effect makes a more favourable prediction of the trapping probability, and is still within an order of magnitude of the number of events observed.

\section{Conclusion}
ALPHA has conducted a search for trapped antihydrogen by attempting to identify the annihilation of a synthesised antihydrogen atom as it is released from our magnetic minimum trap.
The diagnostics incorporated into the ALPHA apparatus allow us to determine and control the experimental parameters to a high level of precision and permit us to estimate the number of trapped antihydrogen atoms the experiment would be expected to produce.
For the first time, we have carried out a trapping attempt in which the experimental inputs have been well-determined and theoretical estimates predict a good probability of detecting trapped antihydrogen.
In this experiment, we have identified six events that are excluded as particles from cosmic rays and, based on simulations, have a very low probability of being due to the release of bare antiprotons which can be trapped in the magnetic minimum trap.
However, without further investigation, we can not definitively claim that these events correspond to trapped antihydrogen.
A higher rate of observed events would greatly facilitate study and characterisation, and presently, reducing the temperature of the component plasmas seems to offer the most promise to this end.
The experimental tools and simulation techniques developed for this trapping attempt and discussed in this paper will be of great utility in our further searches.

\section{Acknowledgements}
This work was supported by CNPq, FINEP/RENAFAE (Brazil), ISF (Israel), MEXT (Japan), FNU (Denmark), VR (Sweden), NSERC, NRC/TRIUMF, AIF (Canada), DOE, NSF (USA), EPSRC and the Leverhulme Trust (UK).
We are also grateful to the AD team for the delivery of a high-quality antiproton beam, and to CERN for its technical support.
We also acknowledge the work of D. Seddon, J. Thornhill and D. Wells (University of Liverpool) on the construction of the vertex detector.

\bibliographystyle{elsart-num}

\end{document}